# Mode Mismatch Mitigation in Gaussian-Modulated CV-QKD

Svitlana Matsenko[(1)], Amirhossein Ghazisaeidi[(2)], Marcin Jarzyna[(3)], Mikkel N. Schmidt[(4)], Søren F. Nielsen[(4)], Konrad Banaszek[(3,5)], Darko Zibar[(1)]

[(1)] DTU Electro, Technical University of Denmark, DK-2800, Kgs. Lyngby, Denmark, svitma@dtu.dk
[(2)] Nokia Bell Labs, 91300 Massy, France,
[(3)] Centre for Quantum Optical Technologies, CeNT, University of Warsaw, 02-097 Warszawa, Poland,
[(4)] DTU Compute, Technical University of Denmark, DK-2800, Kgs. Lyngby, Denmark
[(5)] Faculty of Physics, University of Warsaw, Pasteura 5, 02-093 Warsaw, Poland

**Abstract** *Technical limitations in pulse shaping lead to mode mismatch, which significantly reduces the secure key rate in CV-QKD systems. To address this, a machine learning approach is employed to optimize the transmitter pulse-shape, effectively minimizing mode mismatch and yielding substantial performance improvements.* ©2025 The Author(s)

**Introduction**

Continuous-variable quantum key distribution (CV-QKD) is being investigated as a promising solution for enabling secure communication over current and future fiber-optic networks[1]. Unlike other quantum key distribution approaches, CV-QKD can be implemented using existing coherent transceiver technology and demonstrates strong resilience to coexistence with wavelength-division multiplexing (WDM) traffic.

In recent years, numerous laboratory demonstrations and field trials of various CV-QKD systems have been reported[2]-[6]. However, many of these theoretical studies and experimental implementations assume access to substantial computational resources and idealized operating conditions, particularly for digital signal processing (DSP) at both the transmitter and receiver sides. Transitioning to real-time implementation and commercialization of CV-QKD requires careful consideration of the complexity of digital signal processing (DSP) at both the transmitter and receiver, as well as its performance under realistic operating conditions.

One important practical issue is the *finite length* (finite number of taps) of the transmitter-pulse-shaping and the receiver-filter in practical implementations. When the transmitter and receiver filters are implemented as ideal*, infinite-length* root-raised cosine (RRC) filter pairs, they form a complete orthonormal set of modes[7]. This structure aligns with the requirements of the theoretical security proofs for CV-QKD.

In practice, implementing *infinite-length* or very large root-raised cosine (RRC) pulse-shaping and receiver filters is not feasible. Longer filters significantly increase computational complexity and power consumption. To meet practical constraints, the pulse shaper and the receiver filter must be truncated (their length must be limited) to reduce complexity. However, this truncation introduces a mode mismatch between the transmitter and receiver, leading to intra-symbol interference (ISI) and contributing to excess noise. In our previous work[7], we demonstrated the degradation in secure key rate (SKR) caused by mode mismatch resulting from the truncation of the transmitter's root-raised cosine (RRC) pulse-shaping filter. The penalty arises because, once the number of pulse-shaping filter taps is constrained to a finite value, the transmitter and receiver RRC filter pair becomes suboptimal, leading to mode mismatch as previously discussed. Determining an optimal transmitter pulse-shaping filter that minimizes mode mismatch under a finite-length constrain, while assuming a fixed RRC filter at the receiver, is a complex optimization problem.

In this paper, we propose a machine learning-based approach using the reinforcement learning algorithm[8] to address the aforementioned optimization problem in Gaussian-modulated CV-QKD systems. To the best of our knowledge, this is the first numerical demonstration showing that transmitter–receiver mode mismatch can be significantly reduced, resulting in a substantial improvement in the secure key rate.

**Theoretical Framework**

We assume that Alice transmits a Gaussian-modulated CV-QKD signal $E^{\text{in}}(t)$ to Bob, expressed as:

$$E^{\text{in}}(t) = \sum_{j=-\infty}^{+\infty} \alpha_j^{\text{in}} u(t - jT), \qquad (1)$$

where $\alpha_j^{\text{in}}$ is the Gaussian complex symbol, $u(t)$ is the transmitted pulse shape and $T$ is the symbol duration. Under ideal conditions, $u(t)$ has an infinite length, while for practical applications, it needs to be limited to a couple of symbol periods $T$. Limiting $u(t)$ will imply that different time slots of $T$ duration may not necessarily be orthogonal. After propagating through the quantum channel, the corresponding received symbol is $\alpha_j^{\text{out}} = \sqrt{\tau_{\text{ch}}} \alpha_j^{\text{in}} + n_j$, where $\tau_{\text{ch}}$ is the channel transmittance, and $n_j$ is the complex Gaussian random variable representing the quantum noise, with zero mean and variance equal to $n_{ch}$ in photon number units. The matched receiver filter's impulse response at Bob's side is denoted by $v(t)$. The resulting signal after receiver matched filtering is expressed as:

$$c_j = \int_{-\infty}^{+\infty} dt\, v^*(t-jT) u(t) \cdot \quad (2)$$

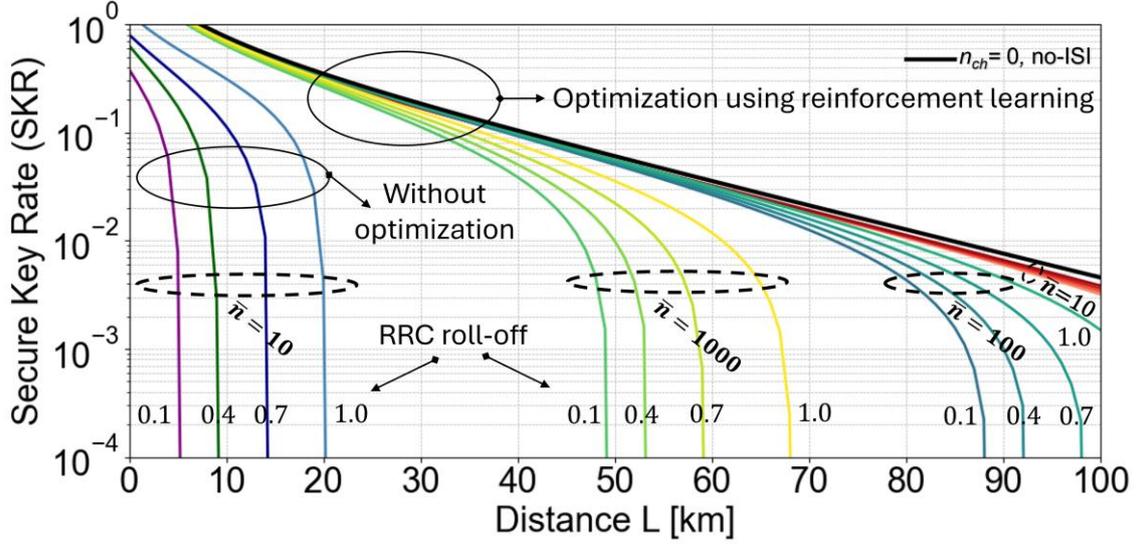

**Fig. 1:** Secure Key Rate (SKR) as a function of transmission distance for different roll-off factors, signal mean photon number, $\bar{n}$, assuming samples-per-symbol equal to 4. RRC roll-off refers to the roll-off factor $\alpha$.

If $u(t) = v(t)$ and both pulse shapes are ideal RRC, then $c_j = \delta_{j0}$ which leads to no ISI. However, due to practical constraints $u(t) \neq v(t)$ which results in a transmitter-receiver mode mismatch and $c_j \neq \delta_{j0}$. In this case, the mode mismatch will result in ISI terms, which will contribute to the excess noise $n_{ex}$ as follows

$$n_{ex} = n_{ch} + \tau_{ch}\bar{n}\sum_{j\neq 0}|c_j|^2, \quad (3)$$

where $\bar{n}$ is the mean photon number of the transmitted signal. Using Eq. (3) for the excess noise, the secret key rate (SKR) for the Gaussian modulated CV-QKD is given by [9], [10]

$$SKR = \beta \log_2\left(1+\frac{\tau_{ch}|c_0|^2\bar{n}}{n_{ex}+1}\right) - g(v_+) - g(v_-) + g(v), \quad (4)$$

where $g(x) = (x+1)\log_2(x+1) - x\log_2 x$, $\beta$ is the reconciliation efficiency and

$$v_\pm = \frac{1}{2}\left\{\sqrt{[(1-\tau)\bar{n}+n_{ex}+1]^2 + 4\tau\bar{n}n_{ex}} \pm [(1-\tau)\bar{n}-n_{ex}]-1\right\} \quad (5)$$

$$v = \frac{1-\tau+n_{ex}}{\tau\bar{n}+n_{ex}+1}\bar{n}, \quad (6)$$

where $\tau = \tau_{ch}|c_0|^2$.

The optimization problem is formulated as determining the finite impulse response (FIR) filter taps of the pulse shaper, generating $u(t)$, that maximize the secure key rate (SKR), subject to the constraint of a finite number of taps. The receiver matched filter is non-adjustable and is fixed to be an RRC with a finite number of taps. The aforementioned problem can be solved using classical gradient-based optimizers, where Eq. (4) is the objective function to maximize. The optimization variables are the pulse shaper filter taps, $[u_0,\ldots,u_{L-1}]$, where $L$ is the truncation length. Without optimization, as e.g. in our previous work[7], $u_k = u(k\Delta t)$, where $\Delta t = T/s$, $T$ is symbol duration, $s$ is the number of samples-per-symbol, and $[u_0,\ldots,u_{L-1}]$ is the RRC impulse response. In this work, by using a gradient-based optimizer as well as a reinforcement learning (RL) approach[8], we find $[u_0,\ldots,u_{L-1}]$ that maximizes SKR given by Eq. (4). While both approaches give the same results in the current setting, the model-free based reinforcement learning approach is more promising, as we can include extra non-differentiable penalty models in future. Here we only report on RL results.

**Results**
Fig. 1 illustrates the secure key rate (SKR) as a function of transmission distance for various roll-off factors ($\alpha$), mean signal photon numbers ($\bar{n}$), and fiber attenuation is 0.2 dB/km.

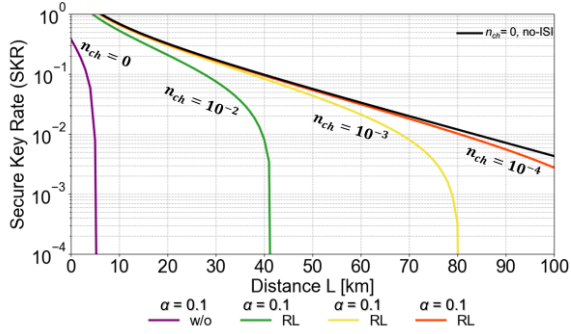

**Fig. 2:** Secret key rate (SKR) and the as a function of distance for channel excess noise $n_{ch}$ = $10^{-2}$, $10^{-3}$, $10^{-4}$ and the roll-off factor $\alpha$ = 0.1, mean photon number $\bar{n}$ = 10. The purple line is without optimization.

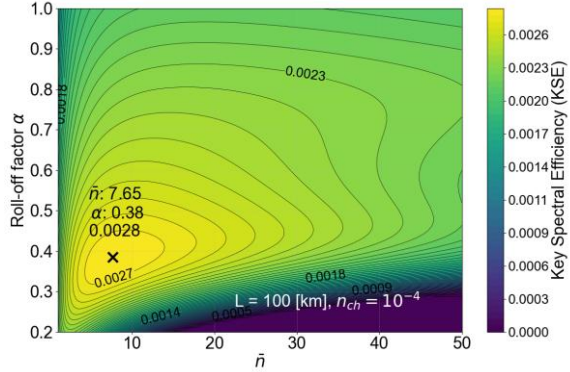

**Fig. 3:** Key spectral efficiency (KSE) as a function of the mean photon number per symbol $\bar{n}$ and the roll-off factor $\alpha$ with RL optimization.

To isolate the impact of mode mismatch, we assume that excess noise is dominated by the mismatch itself rather than the channel. A theoretical upper bound (solid black curve) is shown for the ideal case with zero excess noise. In the simulations, the transmitter pulse-shaping filter consists of 13 trainable taps, while the receiver employs a fixed root-raised cosine (RRC) filter with 101 taps. For the case where the average signal photon number is $\bar{n} = 10$, a substantial performance penalty is observed when the transmitter pulse-shaping filter is fixed to a standard RRC filter (w/o curves), even when there is no excess noise contributed by the channel. This highlights the significant degradation due to mode mismatch. However, when the transmitter filter is optimized using a reinforcement learning (RL curves) approach, the mode mismatch is effectively mitigated. As a result, the SKR of the optimized system shows a negligible penalty compared to the theoretical curve of zero excess noise.

By increasing the signal photon number to $\bar{n} = 100$, and performing the optimization, we observe very little penalty up to 60 km. The penalty then increases depending on the roll-off factor. This is because the mode mismatch is strongly dependent on the signal mean photon number $\bar{n}$, as shown in[7], making it more challenging to compensate for mode mismatch. For $\bar{n} = 100$ without performing the optimization, the SKR was very low (close to zero). We have chosen not to plot it in order to keep Fig. 1 uncluttered.

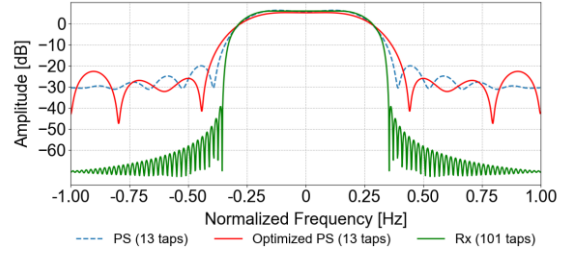

**Fig. 4:** Amplitude response of the pulse-shaping and receiver filter as a function of normalized frequency.

Next, the mean signal photon number $\bar{n}$ is increased to 1000, and the optimization using the RL is performed. It is observed that the penalty increases. However, without the optimization the SKR was almost zero. Next, we investigate if the mode mismatch optimization is still effective if the channel excess noise $n_{ch}$ is present in the system. In Fig. 2, SKR is plotted as a function of the transmission distance for channel excess noise $n_{ch}$=$10^{-2}$, $10^{-3}$ and $10^{-4}$, and mean photon number $\bar{n} = 10$. For low values of channel excess noise $n_{ch}$=$10^{-4}$, it is observed that there is very little penalty. This can be fully attributed to the channel excess noise, which cannot be compensated. As expected, increasing the channel excess noise results in an increased penalty, which can be contributed to the channel excess noise.

Fig.3 shows the key spectral efficiency (KSE), KSE = SKR/(1 + $\alpha$), with RL optimization, which takes into account the bandwidth overhead resulting from a non-zero value of the roll-off factor $\alpha$. The results can be used to find the optimum mean photon number $\bar{n}$ and the roll-off factor $\alpha$.

In Fig. 4, we plot the amplitude response of the optimized pulse-shaper filter together with an RRC filter with 13 and 101 taps, respectively. It is observed that the amplitude response of the optimized pulse-shaping filter is slightly more rounded compared to the RRC. In addition, it also occupies slightly more bandwidth.

**Conclusions**

Practical constraints of the pulse-shaping filter result in a mode mismatch, which induces a penalty in the SKR. We have numerically shown that the mode-mismatch can be effectively mitigated, even in the presence of channel excess noise, and for various mean photon numbers. For practical evaluation in the near future, the learned pulse-shaping filter will be applied in an experimental environment.

**Acknowledgements**

M.J. and K.B. acknowledge support by the European Union's Horizon Europe research and innovation programme under the project 'Quantum Security Networks Partnership' (QSNP, Grant Agreement No. 101114043) and the 'Quantum Optical Technologies' project (FENG.02.01-IP.05-0017/23) carried out within the International Research Agendas programme of the Foundation for Polish Science co-financed by the European Union under the European Funds for Smart Economy 2021-2027 (FENG). D. Zibar and S. Matsenko are supported in part by VILLUM FONDEN by Grant VI-POPCOM and MARBLE (VIL5448 and VIL40555)